\documentclass{appolb}
\usepackage{epsfig}

\begin{document}

\def\lp{\left(}   
\def\rp{\right)}    

\def\SF{\Sigma_{fr}}

\title{Particle Freeze-out within 
 the Self-Consistent Hydrodynamics%
\thanks{Presented at   the 
IV International Workshop on Particle Correlations and  Femtoscopy,  Cracow, September 11-14, 2008.
}%
}
\author{ K.A. Bugaev
\address{Bogolyubov Institute for Theoretical Physics,  
Kiev, Ukraine}
}
\maketitle

\vspace*{-0.5cm}

\begin{abstract}
Here I  discuss some implicit  assumptions of modern hydrodynamic models and argue that
their accuracy cannot be better than 10-15 \%. 
Then I  formulate the correct conservation laws for the fluid emitting particles from an the arbitrary freeze-out (FO) hypersurface (HS)  and show that the derived momentum distribution function  of emitted particles 
does not contain negative contributions which appear in the famous Cooper-Frye 
formula. 
Further I analyze the typical pitfalls of some hydro models trying to  alternatively resolve   the FO problem. 
\end{abstract}
  
\vspace*{0.cm}
  
%
{\bf 1. Introduction.} Relativistic hydrodynamics is one of   the most powerful theoretical tools to study
the dynamics of phase transitions in nucleus nucleus  collisions at 
high energies. 
During last 20 years  it was successfully used 
to model  the phase transition
between  the quark gluon plasma (QGP) and hadronic matter \cite{Rischke1,GYULASSY:01}.
So far,  only within  hydro approach and hydro inspired models  it was possible to find the three major 
signals of the deconfinement transition seen at SPS energies, i.e. the Kink \cite{Kink},
the Strangeness Horn \cite{Horn} and the Step \cite{Step}. 
Nevertheless, from its birth the hydro  modeling of relativistic heavy ion collisions 
suffers from a few severe difficulties which I discuss  in this work along with the self-consistent  formulation of  relativistic hydro equations.

{\bf 2. Explicit and implicit hydro assumptions.}
Relativistic hydrodynamics  is a set of
partial differential equations which describe the local 
energy-momentum and charge conservation  \cite{LANDAU:53} 
%
\begin{eqnarray}
\label{one}
\partial_\mu T^{\mu\nu}_{f} ( x,t) & = & 0\,\,,  \quad \quad T^{\mu\nu}_{f} ( x,t) =   \lp \epsilon_f + p_f \rp 
u_f^\mu u_f^\nu - p_f g^{\mu\nu}\,\,, \\
\label{two}
\partial_\mu N^{\mu}_{f} ( x,t) & = & 0\,\,,  \quad \quad N^{\nu}_{f} ( x,t)  =  n_{f} u_f^\nu \,\,.
\end{eqnarray}
%
Here the components of the energy-momentum tensor $T^{\mu\nu}_{f}$ of the perfect fluid
and its (baryonic) charge 4-current $ N^{\mu}_{f} $ are given in terms of energy density $\epsilon_f$,
pressure $p_f$, charge density $n_f$ and 4-velocity of the fluid $u_f^\nu $.
This is a simple indication  that hydrodynamic description directly probes
the equation of state of the matter under investigation.

As usual to complete the system (\ref{one}) and (\ref{two}) it is necessary to provide

{\bf (A)}  {\it the initial conditions } at some hypersurface and 

{\bf (B)} {\it equation of state} (EOS).

The tremendous complexity  of   {\bf (A)} and {\bf (B)}  transformed each of them into a specialized direction of 
research of  relativistic  heavy ion community.
However, there are several specific features of relativistic hydrodynamics which have to be mentioned.
In contrast to  nonrelativistic hydrodynamics which is an exact science,
the relativistic one, while applied to collisions of hadrons or/and heavy nuclei,  faces  a few problems from the very beginning.  Since the system created during the collision process is small and short living  there were always 
the questions whether the hydro description is good and accurate,  and whether the created  system   thermalizes 
sufficiently fast in order that hydro description can be used. 

Clearly, these  two questions cannot be answered within the framework   of hydrodynamics.
One has to study these problems in a wider frame, and there was some progress achieved
on this way.  
However, there are several implicit assumptions which are difficult to verify 
for the heavy ion collisions (HIC). Thus, we implicitly assume that the EOS of infinite system may successfully 
describe the phase transformations in a finite system created in collisions. 
The exact solutions of several  statistical models both with a  phase transition \cite{PTfiniteV:1} and without  it \cite{PTfiniteV:2}   found for finite volumes teach us that in this case the analog of mixed phase consist of several metastable states which may transform into each other. Clearly,  such a  process cannot be described by the usual hydrody which is dealing with the stable states. 

Furthermore,  usually  it is implicitly assumed that the matter created during the HIC is  homogeneous. 
However, the realistic  statistical models of strongly interacting matter  \cite{Bugaev:07,CGreiner:06} tell us    that 
at and above the cross-over  this  matter consists of  QGP bags with the mean volume of several cubic fm. 
Moreover, the model of QGP bags with surface tension  \cite{Bugaev:07}  predicts an existence of very complicated 
shapes of such bags  above the cross-over due to negative surface tension. 
Note that  the existence  of QGP bags of  such a volume is supported by the model of QGP droplets  \cite{Droplets}
which successfully  resolved the HBT puzzles at RHIC.
 
Also the assumption that the heavy QGP bags  (resonances) are stable compared to the typical life-time of 
the matter created in the HIC is, perhaps, too strong.  The recent results obtained  within the finite width model 
\cite{FWM}  show that  in a vacuum  the mean  width of a resonance of mass $M$   behaves as $\Gamma (M) \approx 600 \left[\frac{M}{M_0}\right]^{\frac{1}{2}}$ MeV (with $M_0 \approx 2$ GeV), whereas in a media it grows with the temperature.   
At the moment it is unclear how the finite width of  QGP bags and other implicit assumptions affect the accuracy 
of hydrodynamic simulations, but  from the discussion above  it is clear that  their a priori accuracy  cannot be better than 10-15 \% \cite{BD:00,SHUR:01}. 
In fact, from the hydro estimates of the HBT radii at RHIC one concludes that, depending on the model, 
the real accuracy could be between 30 \% to 50 \%. 
Clearly,  the same  is true for the hydro-cascade \cite{BD:00,SHUR:01} and  hydro-kinetic \cite{SIN:02} 
approaches. Thus, at present  there are no strong reasons  to believe that  these approaches  are qualitatively 
better than the usual hydrodynamics.

{\bf 3.  Boundary conditions.}
In addition to the assumptions discussed  above,  to complete  relativistic hydrodynamics 
it is necessary to know
 {\it the boundary conditions} which must be consistent with the conservation laws (\ref{one}) and 
(\ref{two}). 
The latter   is known as the {\it freeze-out problem}, and 
it has   two basic aspects \cite{LANDAU:53}: 
{\bf (C1)} the hydro equations should be terminated at the FOHS
$\SF( x,t)$ beyond which the hydro description is not valid;
{\bf (C2)} at the FOHS $\SF( x,t)$ all interacting  particles should be converted into the free-streaming
particles which go into detector without collisions.

The complications come from the fact that the FOHS cannot be found a priory without 
solving  the hydro equations (\ref{one}) and (\ref{two}).  This is a consequence  of relativistic causality on the time-like (t.l.)  parts of the FOHS.\footnote{In  this work I analyze  the two dimensional hydro to which the four dimensional one 
can be always reduced. Then the t.l. HS is defined by the positive element square $ds^2=dt^2 -dx^2 >0$, whereas the space-like HS is   defined by $ds^2 < 0$.}

Therefore, the {\it freeze-out criterion} is usually formulated
as an additional equation (constraint) $F(x,t^*) = 0$ with the solution  $  t =t^*(x)$ which has to be 
inserted   into  the conservation laws and
solved simultaneously with them.

There were many unsuccessful  attempts  to resolve this problem (for their incomplete  list 
see \cite{BUG:96})  by a priory imposing 
the form of  the FOHS, but all of them led to severe difficulties - 
either to negative number of particles or break up of  conservation laws.
The major difficulty  is that the hydro equations should be terminated in such a way, 
that their solution remains unmodified by this very fact.  In addition, this problem 
cannot be postponed to later times because at the boundary with vacuum the particles
start to evaporate from the very beginning of hydro expansion, and this fact should be 
accounted by equations as well. 

The hydrodynamic solution of the FO problem was found in \cite{BUG:96}  and developed further in 
\cite{BUG:99a}. 
This problem  was solved  after a realization of a  fact that at  the t.l.  parts of  the FOHS
there is a fundamental difference between the particles of fluid and the particles
emitted  from its surface: the  EOS of the fluid can be anything, but it implies a zero
value of the mean free path, whereas, according to Landau \cite{LANDAU:53},   the emitted particles cannot  interact at all because
they have an infinite mean free path. Therefore, it was necessary to extend the  conservation laws (\ref{one}) and 
(\ref{two}) from a fluid alone to a  system 
consisting of a  fluid and  the particles of gas emitted  (gas of free particle) from the FOHS. 
The resulting  energy-momentum tensor  and baryonic current (for a single particle species)  
of the  system 
can be, respectively,  cast as 
\vspace*{-0.05cm}
\begin{eqnarray}
\label{three}
T^{\mu\nu}_{tot} ( x,t) & = & \Theta_f^*~T^{\mu\nu}_{f} ( x,t) +  ~\Theta_g^*~T^{\mu\nu}_{g} ( x,t)\,, \\
\label{four}
N^{\mu}_{tot} ( x,t) & = & \Theta_f^*~N^{\mu}_{f} ( x,t)~ +  ~\Theta_g^*~N^{\mu}_{g} ( x,t)\,, 
\end{eqnarray}

\vspace*{-0.05cm}

\noindent
where at the FOHS the energy-momentum tensor of the gas $T^{\mu\nu}_{g} $
and its baryonic current $N^\mu_g$  are  given in terms of {\it the cut-off distribution function} \cite{BUG:96} of particles that have the 4-momentum $p^\mu$
\vspace*{-0.05cm}
\begin{eqnarray}
\label{five}
\phi_g & = & \phi_{eq} \left( x, t^*,  p \right) \,  \Theta\left(p_\rho d \sigma^\rho \right)\,\, ,\\
\label{sixA}
T^{\mu\nu}_{g}\left( x, t^* \right)  & = & \int \frac{d^3 { p}}{p_0} \, p^\mu p^\nu \, \phi_{eq}\left( x, t^*, p \right)    
\Theta \left ( p^\mu d \sigma_\mu \right) \,\,, \\
\label{sevenA}
N^{\mu}_{g}\left( x, t^* \right)  & = & \int \frac{d^3 { p}}{p_0} \, p^\mu ~ \, \phi_{eq}\left( x, t^*, p \right)    
\Theta \left ( p^\mu d \sigma_\mu \right) \,\,.
\end{eqnarray}
\vspace*{-0.25cm}

\noindent
Here  $\phi_{eq}\left( x, t^*, p \right)    $ denotes the equilibrium distribution function of particles
and $d \sigma_\mu$ are the components of the external normal 4-vector to the FOHS  $\SF( x,t^*)$ 
\cite{ BUG:96,BUG:99a}. 

The important feature of equations (\ref{three})-(\ref{five}) is the  presence of several  $\Theta$-functions. 
The $\Theta_g^* = \Theta(F(x,t)) $ function of the gas
and $\Theta_f^* = 1 - \Theta_g^* $ function of the fluid   can be  explicitly expressed in
terms of the FO criterion and can automatically ensure that the energy-momentum tensor of the 
gas (liquid) is not vanishing  only in the domain where the gas (liquid) exists.
On the other hand 
 $\Theta\left ( p^\mu d \sigma_\mu \right) $ function ensures that only the outgoing particles leave 
the fluid domain and go to the detector. Such a  form of the distribution function (\ref{five}) not only
resolves the negative particles  paradox of the famous Cooper-Frye formula \cite{COOP:74}  at
the t.l.  parts of the FOHS, but it allows one to express the hydrodynamic quantities of the gas
of free particles in terms of the invariant momentum spectrum measured by detector.
I would like to stress that  the cut-off distribution (\ref{five}) was {\it rigorously derived}
\cite{BUG:96} within the simple kinetic model, suggested in \cite{gorSin}.

{\bf 4.  The self-consisten  hydro equations.}
The analysis of Refs.   \cite{ BUG:96,BUG:99a}  shows that 
the equations of motion for  the full system 
\begin{equation}\label{eight}
\partial_\mu T^{\mu\nu}_{tot} (x, t)  = 0\,, \quad  \partial_\mu N^{\mu}_{tot} (x, t) = 0 
\end{equation}
are   split  into two subsystems
\begin{eqnarray}
\label{six}
\hspace*{-0.5cm}
 \Theta_f^*~\partial_\mu T^{\mu\nu}_{f} ( x,t) & = & 0\,\,,   \hspace*{1.81cm} \quad 
  \Theta_f^*~\partial_\mu N^{\mu}_{f} ( x,t) ~ = ~ 0\,\,,    
 \end{eqnarray}
\begin{eqnarray} 
\label{seven}
\hspace*{-0.5cm}
d \sigma_\mu   T^{\mu\nu}_{f}  ( x,t^*) & = & d \sigma_\mu T^{\mu\nu}_{g}  ( x,t^*) \,\,, \quad 
d \sigma_\mu   N^{\mu}_{f}  ( x,t^*) ~ =  ~  d \sigma_\mu N^{\mu}_{g}  ( x,t^*) \,\,,
\end{eqnarray}

\vspace*{-0.25cm}
\noindent
since equations for the gas of free particles, $\partial_\mu T^{\mu\nu}_{g}   \equiv 0$ and 
$\partial_\mu N^{\mu}_{g}  \equiv 0$, are identities due the fact that the trajectories  of free particles 
are straight lines. 

Here Eqs. (\ref{six}) are the equations of motion of the fluid, whereas  Eqs. (\ref{seven}) are the boundary
conditions for the liquid at the FOHS.  On the other hand (\ref{seven}) is a system of the nonlinear partial differential equations to find the 
FOHS   $\SF(x,t^*)$ for a given FO criterion.
To  find the FOHS $\SF(x,t^*)$ the solution of the fluid equations (\ref{six})
should be used as an input  for (\ref{seven}). 

There is  a fundamental difference between the equations of motion (\ref{one})  of traditional hydrodynamics  and the corresponding equations (\ref{six}) of hydrodynamics with particle emission:
if the FOHS is found, then, in contrast to the usual hydrodynamics,  the equations (\ref{six}) automatically
vanish in the domain where the  fluid is absent. In this way  the equations  (\ref{three})-(\ref{seven}) resolve
the FO problem in relativistic hydrodynamics. 

In addition, as shown in  \cite{BUG:99a} 
for a wide class of hadronic EOS
these equations resolve the usual  paradox of relativistic  hydrodynamics of finite systems
which is known as {\it a recoil problem} due to  the  emission of particles. The latter means
that a substantial  emission of particles from the t.l. parts of the FOHS 
 is expected to inevitably modify the hydrodynamic solution interior  the fluid. 
However, 
this is not the case for a wide class of realistic EOS of hadronic matter because
at  the t.l.  parts of the FOHS  there appears a new kind of hydro discontinuity, 
{\it the freeze-out shock} \cite{BUG:96}.  The FO shock  is a generalization of
the usual hydrodynamic shock waves  \cite{BUG:87,BUG:88} which for the nonrelativistic flows transforms into the usual hydrodynamic shock.  As shown in \cite{BUG:99a} the supersonic FO shock
is not only thermodynamically stable, i.e. in such a shock the entropy increases,  but also   it 
propagates interior the  fluid faster than the information about the possible 
change of hydrodynamic solution. 

{\bf 5.  Concluding remarks.}
The hydrodynamic  solution of the FO problem required an insertion of the boundary conditions into the conservation laws for the fluid and emitted particles. The subsequent  transport simulations \cite{BRAVINA:99}  showed that  the assumptions of thermal equilibrium at the FOHS
and  small width of the FO front at the t.l. parts of the FOHS are quite  reasonable, whereas the main problem 
appears at the s.l. FOHS where the decay of shortly living  resonances may essentially modify the equilibrium 
distribution function. This problem, however, requires more complicated hydro-kinetic  models \cite{SIN:02} or 
even the kinetic approach with specific boundary conditions  \cite{BUG:02}. 

Further attempts of the Bergen group \cite{LASZLO} to improve the suggested hydro solution of the FO problem were based on 
the hand waiving arguments and, hence, they did not lead to any new discovery.
Note also that from time to time the erroneous attempts  to resolve the FO problem appear \cite{Ivan}, but as usual  they are running into  severe troubles. Thus, in  \cite{Ivan} (and subsequent works) the  artificial $\delta$-like drains in relativistic hydrodynamic equations were inserted, which, besides other pitfalls,  in principle cannot  reproduce  the nonrelativistic hydro equations   even for weak flows.

\vspace*{-0.5cm}

%
%

\end{document}